\documentclass[twocolumn,showpacs,preprintnumbers,amssymb,nofootinbib, aps,prd, eqsecnum]{revtex4-1}
\usepackage{graphicx, booktabs, epsf, epsfig}% Include figure files
\usepackage{bm} % Include bold math: \bm{} creates bold letters and symbols in math mode
\usepackage{amsmath}
\usepackage{mathpazo}

\usepackage{color}
\usepackage{ulem} 
\usepackage{float}

\definecolor{green}{rgb}{0.1,0.6,0.1}

\def\be{\begin{equation}}
\def\ee{\end{equation}}
\def\beq{\begin{eqnarray}}
\def\eeq{\end{eqnarray}}
\def \bsp{\begin{split}}
\def \ensp{ \end{split} }

\def\nn{\nonumber}
\def\f{\frac}

\begin{document}

\title{Black hole spectroscopy for KAGRA future prospect in O5}

\author{Nami Uchikata$^1$}\email{uchikata@astro.sc.niigata-u.ac.jp}

\author{Tatsuya Narikawa$^2$}\email{narikawa@tap.scphys.kyoto-u.ac.jp}

\author{Kazuki Sakai$^3$}\email{k-sakai@nagaoka-ct.ac.jp}

\author{Hirotaka Takahashi$^{4,5}$}\email{hirotaka@kjs.nagaokaut.ac.jp}

\author{Hiroyuki Nakano$^6$}\email{hinakano@law.ryukoku.ac.jp}

\affiliation{
$^{1}$ Graduate School of Science and Technology, 
Niigata University, Niigata 950-2181, Japan\\
$^2$ Department of Physics, Kyoto University, Kyoto 606-8502, Japan \\
$^{3}$Department of Electronic Control Engineering,
National Institute of Technology, Nagaoka College, Niigata 940-8532, Japan\\
$^4$Department of Information and Management Systems Engineering,
Nagaoka University of Technology, Niigata 940-2188, Japan\\
$^5$ Earthquake Research Institute, The University of Tokyo, 
Tokyo 113-0032, Japan\\
$^6$Faculty of Law, Ryukoku University, Kyoto 612-8577, Japan 
}
\date{\today}

\begin{abstract}
Ringdown gravitational waves of compact binary mergers are an important target to test general relativity.
The main components of the ringdown waveform after merger are black hole quasinormal modes.
In general relativity, all multipolar quasinormal modes of a black hole should give the same values of black hole parameters.
Although the observed binary black hole events so far are not significant enough to perform the test with ringdown gravitational waves, it is expected that the test will be achieved in third generation detectors.
The Japanese gravitational wave detector KAGRA, called bKAGRA for the current configuration, has started observation, and discussions for the future upgrade plans have also started.
In this study, we consider which KAGRA upgrade plan is the best to detect the subdominant quasinormal modes of black holes in the aim of testing general relativity.
We use a numerical relativity waveform as injected signals that contains two multipolar modes and analyze each mode by matched filtering.
Our results suggest that the plan FDSQZ, which improves the sensitivity of KAGRA for broad frequency range, is the most suitable configuration for black hole spectroscopy.
\end{abstract}

\pacs{04.70.-s}

\maketitle

\newpage
%%%%%%%%%%%%%%%%%%%%%%%%%%%%%%%%%%%%%%%%%%%%%%%%%%%%%
\section{Introduction}
%%%%%%%%%%%%%%%%%%%%%%%%%%%%%%%%%%%%%%%%%%%%%%%%%%%%%

Detection of gravitational waves from compact binary coalescence by LIGO and Virgo provides a new tool to understand the strong gravitational field~\cite{150914,151226,o1,170104,170814,170817,170608,gwtc-1}.
These gravitational waves are important targets to test general relativity (GR)~\cite{grtest,ligo2}.
The waveform of compact binary coalescences can be divided into three parts; inspiral, merger, and ringdown.
A brief review on the current and future test of GR with gravitational waves is given by Carson and Yagi~\cite{Carson:2019yxq}.
As a useful approach to test GR,
the parameterized post-Einsteinian formalism has been proposed by Yunes and Pretorius~\cite{Yunes:2009ke}.
An inspiral-merger-ringdown consistency test with higher harmonic modes for the LIGO-Virgo's second observation (O2) run is presented by Breschi et al.~\cite{Breschi:2019wki}.
In this study, we focus on the waveform after the merger of two compact objects forming a black hole.

The ringdown gravitational waveform contains black hole quasinormal modes (QNMs), which are characteristic oscillations of the black hole derived from linear perturbations of black hole solutions.
The exact values of black hole QNMs are given when the black hole parameters are given~\cite{leaver} based on GR.
(See Ref.~\cite{berti4} for details on theoretical point of view.)
In other words, QNMs in the ringdown waveform give the information of the remnant black hole, the mass and spin (see, e.g., a fitting formula in Ref.~\cite{echeverria}).
For the binaries with total mass of a few tens of solar mass as observed in O1 and O2, their QNM ranges in a few hundred Hertz, where the sensitivity of ground based detectors are stable~\cite{gwtc-1}.

We can test GR from the QNMs by two ways. 
One is to analyze just the dominant mode of the harmonic $(l,m)=(2,2)$.
Since there is a forbidden QNM parameter region in GR, we can rule out Schwarzschild or Kerr black holes if the observed QNMs are locating in the forbidden parameter region~\cite{nakano2}.
The other one is to analyze higher multipolar $(l,m)$ modes beside the dominant $(l,m)=(2,2)$ mode (where we ignore overtones with $n \neq 0$ in this paper), i.e., ``black hole spectroscopy''~\cite{Detweiler:1980gk}.
In GR, all multipolar modes of a black hole should give the same values of black hole mass and spin within the measurement error range.
This consistency test can put constraint on the other modified gravity theories~\cite{dreyer}.
In this study, we focus on the latter way to test GR from ringdown analysis.

There is a difficult problem for the ringdown gravitational wave data analysis.
We do not know when the QNMs start in the ringdown waveform.
Without knowing the starting time of the QNMs, the result of the analysis is always biased from the theoretically expected values (see, e.g., Fig.~5 in Ref.~\cite{grtest}).
This is because the amplitude becomes maximum at the merger where the nonlinearity is important, before the QNMs start, and for the late time of the waveform, the signal-to-noise ratio (SNR) becomes too small to detect.
There are some studies that attempt to solve this problem.
An empirical method to obtain the starting time of QNM is proposed in Ref.~\cite{Carullo1}.
Some other studies show that the QNMs can be determined from the merger time 
and the estimated parameters become unbiased when the overtone modes are included~\cite{giesler,isi,Bhagwat}.
A method to obtain the final spin irrespective to the starting time of QNM is also proposed~\cite{Ferguson:2019slp}.
Some of the authors of this study also tried to
solve this problem~\cite{sakai}.

Several studies have attempted to analyze the ringdown part of the waveform of binary black hole mergers produced by numerical relativity (NR) simulations.
Berti et al.~\cite{berti1, berti2} analyzed several multipolar modes of the ringdown waveform separately using some fitting techniques in the absence of noise.
They showed the convergence of extracted parameters of the ringdown waveform in the late time of the waveform.
Overtone modes are considered in Ref.~\cite{london}.
Berti et al. also showed the event loss by matched filtering two-mode waveforms by using single mode templates~\cite{berti3}.
In terms of the test of no-hair theorem, Gossan et al.~\cite{gossan} analyzed two-modes damped sinusoidal waveforms by Bayesian analysis and the extended study is given in Ref.~\cite{meidam}. 
In these studies, they choose three ringdown parameters, frequency and damping rate of the dominant mode and frequency of the subdominant mode, to use to test no-hair theorem based on Refs.~\cite{kamaretsos,kamaretsos2}.
They give some possible constraints by deviating one of the parameters above from GR.
Further studies also show the importance of analyzing higher multipolar modes~\cite{Bhagwat2, Shaik} and overtones~\cite{Ota}.
Analysis using a parametrized ringdown waveform for nonspinning binaries is given in Ref.~\cite{richard} and parametrized ringdown waveforms related to specific theories are given in Ref.~\cite{Maselli}.
In the study, they use two-mode damped sinusoidal signals instead of NR waveforms. 
A time-frequency analysis was applied to analyze the ringdown waveform in Ref.~\cite{sakai}.
They also investigated to find the starting time of QNMs.
Comparison of several methods to analyze QNM waveforms is shown in Ref.~\cite{Nakano:2018vay}.
Detail study of multimode ringdown analysis is given in Ref.~\cite{Berti:2005ys}, which aims for the space-based gravitational wave detection.

The Japanese ground based gravitational wave detector KAGRA~\cite{Aso:2013eba} has started observation.
Although the sensitivity is expected to be very low during O3, it will be improved to $\sim$ 20 Mpc in the binary neutron star range in O4, which will start at the beginning of 2021.
For the aim of black hole spectroscopy, however, we need much higher sensitivity, since LIGO and Virgo's sensitivities in O1 and O2 are still not sufficient to analyze black hole QNM (see, e.g., Ref.~\cite{Carullo2}).
Several upgrade configurations are proposed for KAGRA in O5, tentatively called KAGRA plus (KAGRA+)~\cite{K+}.

To investigate which KAGRA+ plan has the best possibility for black hole spectroscopy\footnote{There are also studies on black hole spectroscopy for LIGO detectors and their upgrades~\cite{westerweck,Cabero:2019zyt}.},
we use an NR waveform produced by the Center for Computational Relativity and Gravitation in Rochester Institute of Technology
(RIT)~\cite{rit,Healy:2017psd} and construct a waveform by superposing two modes, $(l,m) = (2,2)$ and $(3,3)$, from the NR simulation.
The $(l,m) = (2,2)$ mode is the dominant mode, and then the $(l,m) = (3,3)$ and $(4,4)$ modes
also have the important contribution.
Since we treat the waveform for an unequal mass binary, 
the $(l,m)=(3,3)$ mode can be excited, and more dominant than the $(l,m)=(4,4)$ mode.
The $l \neq m$ modes become sub-dominant and have complex feature due to mode-mixing~\cite{kelly}.
We also add Gaussian noise generated from each sensitivity configuration of KAGRA+ to produce mock data.

Using matched filtering, we analyze two modes separately and estimate the frequency and quality factor of each mode.
To deal with the unknown starting time of QNMs, we set the starting time of the template $t_0$ as a free parameter.
In principle, the frequency and quality factor should converge at large $t_0$.
Then, we evaluate the consistency of estimated black hole parameters between two modes by assuming GR.
Here, we use a single mode damped sinusoidal as a template and change the frequency range for the matched filter integration when analyzing the subdominant mode.
This method makes us possible to detect the subdominant mode even it is superposed with the dominant mode, however, it also makes us difficult to detect the mode with higher frequency and lower amplitude compared to the dominant mode.
Therefore, we restrict ourselves to analyze only two modes; $(l,m)=(2,2)$ and $(3,3)$.

Our purpose is to show a possible method to analyze the ringdown waveform in detail after the compact binary mergers are detected and how further we can analyze using just a single mode damped sinusoidal template.
In our method, we do not assume any information on the ringdown waveform a priori, such as the starting time of QNMs, and relative amplitudes and phases of the subdominant mode.
We only need to refer the merging time of binaries.

This paper is organized as follows. 
In Sec.~II, we summarize the proposed KAGRA future upgrade plans, and describe the method of analysis used in this study and how we analyze two modes.
Also, we explain the properties of a NR waveform that we use to analyze.
In Sec.~III, we show the results of error regions of the frequency and the quality factor for two mode analysis.
We also show the error regions of black hole parameters converted from the estimated parameters.
We summarize the study in Sec.~IV. 
Some additional analyses, including a non Kerr black hole spectroscopy, are presented in Appendices. 

%%%%%%%%%%%%%%%%%%%%%%%%%%%%%%%%%%%%%%%%%%%%%%%%%%%%%%%%%%%
\section{Method of analysis} 
%%%%%%%%%%%%%%%%%%%%%%%%%%%%%%%%%%%%%%%%%%%%%%%%%%%%%%%%%%%
%%%%%%%%%%%%%%%%%%%%%%%%%%%%%%%%%%%%%%%%%%%%%%%%%%%%%%%%%%%
\subsection{Sensitivity curves for KAGRA upgrade plan; KAGRA+} 
%%%%%%%%%%%%%%%%%%%%%%%%%%%%%%%%%%%%%%%%%%%%%%%%%%%%%%%%%%%

So far, four sensitivity curves are proposed for KAGRA future upgrade plan, KAGRA+; LF, 40kg, HF, and FDSQZ~\cite{K+}.
These four plans target different frequency ranges: 

\begin{enumerate}
\renewcommand{\labelenumi}{(\alph{enumi})}
\item
LF targets a lower frequency range by improving sensitivity around 30 Hz with longer suspension fibers.
\item
40kg targets a middle frequency range by using heavier test masses of 40 kg.
\item
HF targets a higher frequency range over 200 Hz by injecting high powered laser of 3440 W and frequency independent squeezing.
\item
FDSQZ is planned to inject frequency dependent squeezed laser with a 30 m filter cavity for improving a broad frequency range.
\end{enumerate}

In the existing gravitational wave detectors, 
bKAGRA, the current KAGRA configuration, already has important features,
an underground site and cryogenic temperatures
to reduce detector's noise.
Therefore, KAGRA should have different technologies
as described above that are possible in this decade.
Also, as a long term plan,
multiple technologies will be combined.

\begin{figure}[t]
\includegraphics{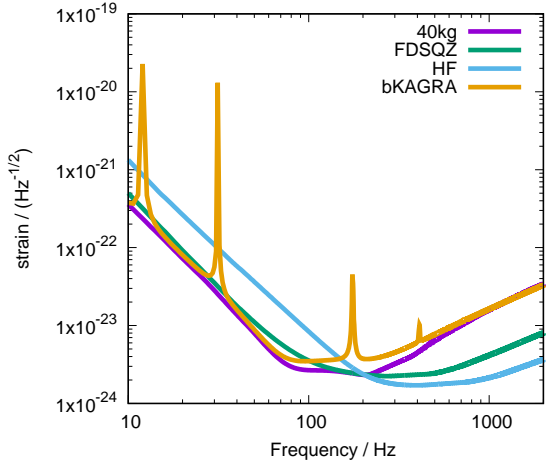}
\caption{Sensitivity curves for three KAGRA+ plans, FDSQZ, HF, 40kg, and bKAGRA.  
Note that line noises are not included yet in the KAGRA+ sensitivities.}
\label{sensitivity}
\end{figure}

In this study, we focus on three of the upgrade plans; HF, FDSQZ, and 40kg, since the frequency range of LF is too narrow and may not appropriate for our aim to analyze multimode spectrum. 
The strain sensitivities of those plans and bKAGRA are shown in Fig.~\ref{sensitivity}.
It is noted that line noises are not included yet in the KAGRA+ sensitivities.

%%%%%%%%%%%%%%%%%%%%%%%%%%%%%%%%%%%%%%%%%%%%%%%%%%%%%%%%%%%
\subsection{Matched filtering and template} 
%%%%%%%%%%%%%%%%%%%%%%%%%%%%%%%%%%%%%%%%%%%%%%%%%%%%%%%%%%%

We use matched filtering method to analyze the waveform.
Matched filtering is an optimal method to detect gravitational wave signals buried in Gaussian noise, if we can predict waveforms.
The SNR is given as 
\be
\rho\equiv (x|s_t)= 4 \,  \mbox{Re}\left ( \int ^{f_{\rm max}}_{f_{\rm min}} \f {\tilde{x}(f) \tilde{s_t}^*(f) }{S_n(f)} df \right ) \,,
\label{snr}
\ee
where $x(t)$ is a detected signal, $s_t(t)$ is a template waveform, and $S_n(f)$ is the noise power spectrum of a detector. 
Tilde denotes the Fourier transform of the corresponding time series function.
The lower and higher cutoff frequencies, $f_{\rm min}$ and $f_{\rm max}$, respectively, depend on the mode of $(l,m)$ and the sampling rate as described below.
In this study, we use an NR waveform added by Gaussian noise instead of a detected signal $x(t)$.
The Gaussian noise is generated randomly from the noise power spectrum for each detector's sensitivity.

%%%%%%%%%%%%%%%%%%%%%%%%%%%%%%%%%%%%%%%%%%%%%%%%%%%%%%%%%%%
%\subsection{Template} 
%%%%%%%%%%%%%%%%%%%%%%%%%%%%%%%%%%%%%%%%%%%%%%%%%%%%%%%%%%%

The QNM waveform with a single mode is described by a damped oscillation 
with a constant frequency, so the template can be simply described by a damped sinusoidal,
\be
s_t(t) = 
\begin{cases}
& 0 \quad (t<t_0) \\
& \f{1}{N} e^{-\pi f_{lm} (t-t_0)/Q_{lm}} \cos [2 \pi f_{lm} (t- t_0) -\phi_0 ] 
\cr & \qquad \qquad \qquad \qquad \qquad \qquad \qquad (t \ge t_0) \,,
\end{cases}
\label{template}
\ee
where $f_{lm}$, $Q_{lm}$, $t_0$ and $\phi_0$ are the frequency, quality factor, 
starting time and initial phase of the template for a given $(l,m)$ mode, respectively.
Here, we have omitted the label of $(l,m)$ in $t_0$ and $\phi_0$ for simplicity.
$N$ is a normalized constant so that $(s_t|s_t) =1$.
The template can be rewritten in the following form,
\be
s_t(t) =\f{1}{N} \left (h_c \cos\phi_0 + h_s \sin \phi_0 \right ) \,,
\ee
where
\be
\begin{split}
h_c &= e^{-\f{\pi f_{lm} (t-t_0)}{Q_{lm}}} \cos [2 \pi f_{lm}(t-t_0)] \,,\\
h_s & =e^{-\f{\pi f_{lm} (t-t_0)}{Q_{lm}}} \sin [2 \pi f_{lm}(t-t_0)] \,.
\end{split}
\ee 
Then, the maximum of the SNR defined by Eq.~\eqref{snr} against the initial phase $\phi_0$ can be obtained as~\cite{tb,tb2},
\be
\rho^2|_{\mbox{max}\, \phi_0}= \f {(x|\hat{h_c})^2+(x| \hat{h_s}) ^2 -2 (x| \hat{h_c})(x|\hat{h_s})(\hat{h_c}|\hat{h_s} )}{1-(\hat{h_c}|\hat{h_s} )^2} \,,
\label{rho}
\ee
where
\be
\begin{split}
\hat{h_c} &= \f{h_c}{\sqrt{(h_c|h_c) }} \,, \\ 
\hat{h_s} &= \f{h_s}{\sqrt{(h_s|h_s) }} \,.
\end{split}
\ee
The phase $\phi_0$ that gives the maximum $\rho$ is given by
\be
\tan \phi_0 = \f{(h_c|h_s)(x|h_c)-(h_c|h_c)(x|h_s)}{(h_c|h_s)(x|h_s)-(h_s|h_s)(x|h_c)} \,.
\ee
Initially, there are four parameters in the template given in Eq.~\eqref{template}; $f_{lm}, Q_{lm}, t_0$ and $\phi_0$.
Using the formula~\eqref{rho}, we can reduce one parameter $\phi_0$.
We assume $t_0$ as a free parameter and search the best fit parameters $(f_{lm},Q_{lm})$ of the template against $t_0$
because the starting time of the template $t_0$, which should coincide with the starting time of QNM in the ringdown waveform, is unknown.

Our interest is to extract each QNM from a waveform of a superposition of two modes.
The details of the waveform is described in the next section.
First, we analyze the waveform by matched filtering using single mode template to find the dominant $(l,m)=(2,2)$ QNM. 
Here, we set $f_{\rm min}=20$ Hz.
Then, to find the subdominant $(l,m)=(3,3)$ QNM,
we cut the frequency lower than the estimated dominant mode frequency, which corresponds to perform a high pass filter. 
In practice, we set the lower frequency of the integral of matched filtering to the estimated dominant mode frequency in Eq.~\eqref{snr}.
Finally, we may estimate the subdominant mode by matched filtering using single mode templates again.

%%%%%%%%%%%%%%%%%%%%%%%%%%%%%%%%%%%%%%%%%%%%%%%%%%%%%%%%%%%
\subsection{Numerical relativity waveform} 
%%%%%%%%%%%%%%%%%%%%%%%%%%%%%%%%%%%%%%%%%%%%%%%%%%%%%%%%%%%

The ringdown part of NR waveforms is expanded by the spin-($-2$) weighted spherical harmonics, $_{-2}Y_{lm}$ as follows~\cite{berti3};
\begin{align}
h(t)  = &\f{1}{r} \sum e^{-\pi f_{lm}t/Q_{lm}} A_{l|m|}  \left [ \cos (2 \pi f_{lm}t -\phi_{lm}) Y^+_{lm}F_+\right. \nonumber \\
& +  \left.  \cos (2 \pi f_{lm}t -\phi_{lm}+\pi/2) Y^{\times}_{lm}F_{\times}\right ] \,,
\end{align}
where
\be
\begin{split}
 Y^+_{lm} & \equiv  {}_{-2}Y_{lm} (\theta,0) + (-1)^l {}_{-2}Y_{lm} (\theta,0) \,,\\
Y^{\times}_{lm} &  \equiv  {}_{-2}Y_{lm} (\theta,0) - (-1)^l {}_{-2}Y_{lm} (\theta,0) \,, 
\end{split}
\ee
and $F_{+, \times}$ are antenna pattern functions of a detector.
It is easy to check that $Y^{\times}_{lm} (\pi/2,0) = 0$.
In this study, we use NR waveforms of the $(l,m) =(2,2)$ and $(3,3)$ modes, and combine them to construct a waveform that contains two modes. 
Since the maximum of $Y^+_{33}/ Y^+_{22}$ is for $\theta = \pi/2$, we choose the inclination angle at $\theta = \pi/2$ so that we can get the maximum effect of the subdominant $(l,m) =(3,3)$ mode.
It should be noted that we do not treat any effect of
spherical-spheroidal mixing~\cite{Buonanno:2006ui}
because of small effect on the modes discussed here.
We also assume an optimal source location, overhead of the detector.

The NR waveform used in this study is RIT:BBH:0317 in a public catalog of black-hole-binary waveforms~\cite{rit,Healy:2017psd}.
The initial condition of binary mass ratio is 0.75.
The spins are aligned with the angular momentum of the orbital plane, i.e., there is no precession, and the dimensionless spin components are both $0.95$.
Estimation values of the ratio of the final mass to total mass and the dimensionless final spin from the simulation are 0.894091 and 0.9425, respectively.

Although matched filtering is the most optimal method for parameter estimation when the waveform is known, the short-lived, exponentially damped waveform is difficult to be analyzed.
Therefore, an NR waveform that has higher final spin is needed, i.e., relatively slow damped QNM.
Also, the subdominant $(l,m)=(3,3)$ mode is much excited when the mass ratio deviates from the unity~\cite{london}.
To analyze the ringdown waveform with the subdominant mode, 
and to investigate which KAGRA+ plan is optimal for black hole spectroscopy,
the above two conditions are needed and RIT:BBH:0317 is an ideal NR waveform for this analysis.

In this study, we rescale the source frame total mass and the redshift to  $(M_{\rm total} \, , \, z)=(70 M_\odot \, , \, 0.0113)$, $(140 M_\odot \, , \, 0.0438)$, and $(280 M_\odot \, , \,0.0852 )$.
The total mass is chosen so that each sensitivity curve we use has an advantage to at least one waveform of the above three cases of scaling.
Here, we focus on two-mode analysis, and in our method we cut off the lower frequency region for the analysis of the (3,3) mode as explained in the previous subsection.
This means that the SNR should be much larger for small total mass cases, since the $f_{22}$ has higher frequency
for these cases.
Therefore, we limit the cases for $M_{\rm total } \ge 70 M_{\odot}$.
Since we are interested in the best performance for the sensitivity curves of the three KAGRA+ plans,
we choose the ideal redshift to imitate golden events.

%%%%%%%%%%%%%%%%%%%%%%%%%%%%%%%%%%%%%%%%%%%%%%%%%%%%%%%%%%%
\section{Results} 
%%%%%%%%%%%%%%%%%%%%%%%%%%%%%%%%%%%%%%%%%%%%%%%%%%%%%%%%%%%
%%%%%%%%%%%%%%%%%%%%%%%%%%%%%%%%%%%%%%%%%%%%%%%%%%%%%%%%%%%
\subsection{Estimation on $(f_{lm}, Q_{lm})$} 
%%%%%%%%%%%%%%%%%%%%%%%%%%%%%%%%%%%%%%%%%%%%%%%%%%%%%%%%%%%

We set the sampling frequency to $8192$ Hz for the $M_{\rm total} = 70 M_{\odot}$ case,  $4096$ Hz for the $M_{\rm total} = 140 M_{\odot}$ case, and  $2048$ Hz for the $M_{\rm total} = 280 M_{\odot}$ case.
We add a random Gaussian noise to the NR waveform and prepare such simulated data for 100 different noise realizations. 

\begin{figure}[t]
\includegraphics{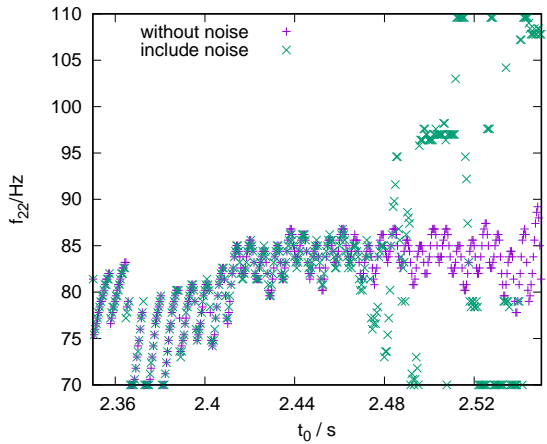}
\caption{A demonstration of estimation of the best fit $f_{22}$
as a function of $t_0$ from the matched filter analysis for $(M_{\rm total}\, , \, z)=(280M_{\odot} \, , 0.0852)$ with KAGRA+FDSQZ.
The horizontal axis represents the starting time of the template.
Each plus (cross) sign represents the analyzed waveform without (with) Gaussian noise.
The peak of SNR is around at 2.4 s.}
\label{sample}
\end{figure}

First, we show how the QNM parameters can be estimated by using our method.
Figure~\ref{sample} shows an example of estimation of $f_{22}$ from matched filtering for one noise realization case 
of the KAGRA+FDSQZ sensitivity curve
compared with the zero noise case.
Each plus/cross sign shows the best fit value at the starting time of the template $t_0$ without/with Gaussian noise. 
The merger time, or the peak of SNR is around at 2.4 s.
We can see that the estimation is less affected by the Gaussian noise until $\sim$ 2.48 s.
It is, however, difficult to estimate at later time where $ \rho \lesssim 7$.

In our analysis, we assume that we already have the information of inspiral and merger parts, that is, we know the merger time and the frequency at the merger in advance.
Since it is difficult to determine the starting time of QNM in the inspiral-merger-ringdown waveform,
we consider the following process to estimate $(f_{lm}, Q_{lm})$.
First, we take a small interval of $t_0$ after the merger time $ t_{\rm merger}$, $t_0^{\rm end} -t_0^{\rm start} (t_0^{\rm start} \ge t_{\rm merger})$.
Then, vary the interval by changing both $(t_0^{\rm start}, t_0^{\rm end})$.
Here, the mean value of $f_{lm}$ of the corresponding interval may change.
Next, we choose the interval where the deviation of $f_{lm}$ from the mean value of $f_{lm}$ becomes minimum and assume the mean value of $f_{lm}$ in that interval as estimated $f_{lm}$. 
For the smaller SNR, the parameter estimation becomes unreliable, so we set thresholds as $\rho_{22} \ge 8$ and $\rho_{33} \ge 5$.
The interval of $t_0$ used in the estimation of the quality factor is the same as that of the frequency of the corresponding multipolar mode,
because the frequency is estimated much better than the quality factor.

Using the method described above, we summarize the symmetric 90\% error regions of estimated parameters for 100 noise realizations for each sensitivity curve
in Table~\ref{result1}.
Rough values of $\rho_{lm}$ at the peak amplitude are also shown in the table.
The distributions of estimated parameters for each sensitivity curve are shown in Appendix~\ref{pdf}.
We should note that the symmetric intervals always exclude the endpoints of distributions, however, the expected value does not locate at endpoints in this study.
Therefore, although the distribution is asymmetric for some cases, the error regions defined by symmetric intervals are sufficient to discuss the accuracy of parameter estimation done in this study.

\begin{table*}[!htb]
 \caption{The symmetric 90\% error regions of $(f_{lm}, Q_{lm})$ for three different KAGRA+ plans and bKAGRA (see Appendix~\ref{pdf} for the distributions of estimated parameters). 
  The frequency is given in the detector frame.
  The numerical values in the parenthesis for $f_{lm}$ and $Q_{lm}$ denote the lower and upper bounds of the error region.
  Rough values of SNR for each $(l,m)$ mode at the peak amplitude, $\rho_{lm}^{\rm peak}$, are also shown. 
  In the Schwarzschild limit, $Q_{22} \sim 2.1$ and $Q_{33} \sim 3.2$.
  Therefore, all cases are consistent with GR in terms of the quality factor in a weak test~\cite{nakano2}.}
\begin{tabular}{l l c c c c c c}
\hline \hline
 $(M_{\rm total}\,,\, z)$ & Plan &  $f_{22}$/Hz& $Q_{22}$ & $f_{33}$/Hz & $Q_{33}$ & $\rho_{22}^{\rm peak}$&  $\rho_{33}^{\rm peak}$\\ 
\hline
 $ (70 M_{\odot}\,,\, 0.0113)$& 40kg & $(374, 390) $& $(5.25, 9.99) $& $(511, 609)$& $(6.74, 16.02 )$& 200& 8
\\
 &  FDSQZ & $(367, 377)$& $(4.95, 7.35) $ & $(519,  597) $& $(4.64, 15.73) $& 250& 20
\\
 & HF & $(367, 374)$  & $(5.00, 7.24)$ & $(542, 591) $& $(4.94, 14.59)$  &300 & 28
\\ 
 & bKAGRA & $(349,383 )$ & $(3.62, 6.42 )$ & $(508, 610) $& $(5.89, 15.99)$ & 130 & 7
\\ 
 & (Expected value)  & 372& 6.57& 576 & 10.09 & &\\ 
%%%%%
 $ (140 M_{\odot}\,,\, 0.0438)$ &40kg & $(175, 181)$ & $(4.88, 6.85)$& $(266, 289)$& $(4.32, 15.08)$  & 170 & 15
\\ 
 &  FDSQZ & $(174, 180)$  & $(5.14, 8.32)$ & $(266,284)$ & $(5.36, 15.46) $  &150 & 17
\\
 & HF & $(170, 178)$ &  $(6.12, 10.10)$ & $(266, 281)$ & $ (5.67, 15.24) $&  105& 19
\\ 
 &  bKAGRA & $(174, 177)$ & $\cdots$ & $(260, 289)$& $(3.76, 14.18)$  & 110 & 10
\\
 & (Expected value)  & 180  & 6.57 & 279 & 10.09& &\\
%%%%%%
 $ (280 M_{\odot}\,,\, 0.0852)$ &40kg &$(83, 87)$ & $(4.47, 8.08)$ & $(123, 136)$ & $ (6.45, 15. 49)$  & 170 & 20
\\ 
 &  FDSQZ & $(83, 86) $& $(6.08, 10.16)$& $(123, 136 )$ & $(7.45, 15.55)$  &100 &18
\\
 & HF &  $(82, 86)$& $(9.39,13.56)$ & $(120,136)$& $(10.98, 16.07)$  &35 & 8
\\ 
 &  bKAGRA & $(83, 87) $ &$( 3.93, 7.69)$ & $(116, 139)$ & $(4.66,15.25) $ &140 & 16
\\
 & (Expected value)  & 87 &6.57 &134& 10.09 & &\\ 
\hline \hline
\end{tabular}
  \label{result1}
\end{table*}

\begin{table*}[!htb]
\caption{The same as Table~\ref{result1} but for the symmetric 68\% error regions.}
\begin{tabular}{l l c c c c c c}
\hline \hline
 $(M_{\rm total}\,,\, z)$ & Plan &  $f_{22}$/Hz& $Q_{22}$ & $f_{33}$/Hz & $Q_{33}$ & $\rho_{22}^{\rm peak}$&  $\rho_{33}^{\rm peak}$\\ 
\hline
 $ (70 M_{\odot}\,,\, 0.0113)$& 40kg & $(376, 384) $& $(5.82, 8.50) $& $(525, 594)$& $(10.85, 16.38 )$& 200& 8
\\
 &  FDSQZ & $(369, 374)$& $(5.30,6.47 ) $ & $(544,  587) $& $(6.51,14.43 ) $& 250& 20
\\
 & HF & $(368, 372)$  & $(5.22,6.44 )$ & $(555, 580) $& $(6.18,12.47 )$  &300 & 28
\\ 
 & bKAGRA & $(355,378 )$ & $(4.05, 5.67 )$ & $(519, 596) $& $(10.81,16.38 )$ & 130 & 7
\\ 
 & (Expected value)  & 372& 6.57& 576 & 10.09 & &\\ 
%%%%%
 $ (140 M_{\odot}\,,\, 0.0438)$ &40kg & $(176, 179)$ & $(5.10,6.05 )$& $(273, 285)$& $(5.76, 12.77)$  & 170 & 15
\\ 
 &  FDSQZ & $(175, 179)$  & $(5.41,6.97 )$ & $(270,281)$ & $(7.16,14.08 ) $  &150 & 17
\\
 & HF & $(172, 176)$ &  $(6.62,8.39 )$ & $(270, 278)$ & $ (6.98,13.61 ) $&  105& 19
\\ 
 &  bKAGRA & $(175, 176)$ & $\cdots$ & $(265, 282)$& $(4.73,11.46 )$  & 110 & 10
\\
 & (Expected value)  & 180  & 6.57 & 279 & 10.09& &\\
%%%%%%
 $ (280 M_{\odot}\,,\, 0.0852)$ &40kg &$(84, 86)$ & $(4.81, 6.53)$ & $(130,135 )$ & $ (6.45, 15. 49)$  & 170 & 20
\\ 
 &  FDSQZ & $(83,85 ) $& $(6.52, 8.70)$& $(128,134  )$ & $(8.63, 14.18)$  &100 &18
\\
 & HF &  $(83, 85)$& $(9.92,12.38)$ & $(123,132)$& $(12.11, 15.47)$  &35 & 8
\\ 
 &  bKAGRA & $(84, 86) $ &$( 4.26, 5.99)$ & $(124,136 )$ & $(6.09,13.47) $ &140 & 16
\\
 & (Expected value)  & 87 &6.57 &134& 10.09 & &\\ 
\hline \hline
\end{tabular}
  \label{result1-2}
\end{table*}

We can see that the frequency is well estimated for any sensitivity curve for both modes.
For massive binary cases, $M_{\rm total} = 140 M_{\odot}$ and $280M_{\odot}$, $f_{22}$ is estimated systematically lower than the expected QNM frequency.
This is because the matched filtering analysis using damped sinusoidal templates is biased by the peak amplitude, where the frequency is lower than the QNM frequency.
For the quality factor, it is found that the expected value is contained in the error region for most of the cases, however, the error region of $Q_{33}$ contains upper end of the search region for some cases.
As we know, the matched filtering is more sensitive
to the gravitational wave phase and its time derivative, i.e., frequency
than the amplitude.
For the $M_{\rm total} = 140 M_{\odot}$ case, the estimation of $Q_{22}$ for bKAGRA is not shown
because the estimated value takes only one value, 
the upper end of the search region.
The conversion of the frequency is slow for bKAGRA compared to other sensitivity curves and the quality factor is not properly estimated when the frequency starts to converge.
The slow conversion of the frequency of bKAGRA might be caused by the line noise near the expected value of $f_{22}$.
Therefore, within our method, valid estimation of the quality factor  cannot be obtained for this case. 

Table~\ref{result1-2}, which presents the symmetric 68\% error regions of estimated parameters,
shows clearer tendencies in the estimation of frequency and quality factor.
The error region of $f_{22}$ is about a half of that of the 90\% error region, for most of the cases.
As for a rough estimate, since the statistical error of the frequency is proportional to $\rho ^{-1}$ from a Fisher matrix calculation~\cite{Berti:2005ys}~\footnote{Note that the statistical error of the frequency is also proportional to the damping time. However, we are comparing the error region within the same mass scale,
therefore we only consider the effect of SNR.}, 
we need about twice louder signals
to have the error regions in Table~\ref{result1-2} as the 90\% regions.
For $f_{22}$ and $Q_{22}$, many cases with sufficiently large SNRs have 
lower frequencies and quality factors than the expected values.
This is quite natural because the overtones ($n \neq 0$) which we have ignored here, 
have lower frequencies and quality factors than those for the $n=0$ mode.
On the other hand, there are reverse cases in KAGRA+40kg and HF.
Since the $(l,m)=(3,3)$ contamination is low (see the values of $\rho_{33}^{\rm peak}$),
these estimations will be unphysical.

In our method, we cut off the frequency region $f \lesssim f_{22}$ for the $(3,3)$ mode estimation, so $\rho_{33}^{\rm peak}$ becomes much smaller even for $\rho_{22}^{\rm peak} \sim O(100)$, especially for the lower mass case.
This means that it is difficult to analyze several multimodes for the lower mass binary system unless the distance between the two frequencies is much closer.  
In our results, $\rho_{33}^{\rm peak}$ for the case of $70M_{\odot}$ with KAGRA+40kg and bKAGRA takes smaller value than the threshold value for some noise realizations.
Therefore, the estimations for 40kg and bKAGRA cases may not be reliable.

As a summary from Tables~\ref{result1} and \ref{result1-2},
we can confirm that KAGRA+FDSQZ performs the best, i.e., all estimations contain expected values except for $f_{22}$ in the $M_{\rm total} = 280M_{\odot}$ case in Table~\ref{result1}.
Although the performance of bKAGRA is lower compared to the results of other KAGRA+, bKAGRA can also give comparable estimations with KAGRA+FDSQZ for the $M_{\rm total} = 280M_{\odot}$ case.

\begin{table*}[htbp]
 \caption{The symmetric 90\% error regions of $(a/M, M/M_\odot)$ for three different KAGRA+ plans and bKAGRA.
  The numerical values in the parenthesis for $(a/M)_{lm}$ and $M_{lm}/M_{\odot}$ denote the lower and upper bounds of the error region.
  The final mass is presented in the detector frame, i.e., 
  the expected value is given by $0.894091 (1+z) M_{\rm total}$.
  The last two columns present the error regions defined by Eq.~\eqref{eq:ER} to see the consistency between two modes.}
\begin{tabular}{l l c c c c c c}
\hline \hline
 $(M_{\rm total}, z)$ & Plan & $(a/M)_{22}$& $M_{22}/M_{\odot}$ & $(a/M)_{33}$ & $M_{33}/M_{\odot}$ & $\delta(a/M)/10^{-2}$ &$ \delta M/M_{\odot} $\\
\hline
 $ (70 M_{\odot}, 0.0113)$& 40kg & $(0.912, 0.977)$& $(58, 70) $ & $(0.844, 0.975)  $& $ (57, 77 )$ &$(-8.1 ,8.3) $ & $(-16, 8)$
\\ 
 &  FDSQZ & $(0.893, 0.954) $ &$(58, 65)$ & $( 0.433, 0.972 )$& $(44, 72) $ & $(-8.3, 42) $ &$(-13, 17)$
\\
 & HF & $ (0.894, 0.953)$ &$( 58, 66 )$ & $(0.659, 0.971)$& $(49,70) $ & $(-6.5 ,26)$ &$(-10,13)$
\\ 
 & bKAGRA & $(0.779, 0.939)$ & $(50, 63)$ & $(0.752, 0.975)$& $(52 , 78)$ & $(-20 , 7.3)$ &$( -24 , 5)$
\\ 
 & (Expected value) & 0.9425 & 63  & 0.9425 & 63 & & \\
%%%%%
 $ (140 M_{\odot}, 0.0438)$ &40kg & $(0.892, 0.947)$ & $(120, 132) $ & $(0.268, 0.965)$& $(83.3, 145)$ & $(-7.5, 61)$ & $(-19, 44)$
\\ 
 &  FDSQZ & $(0.902, 0.965 )$&  $(124, 142) $ & $(0.714, 0.972)$& $(103, 145 )$ &$(-7.1, 21)$ & $(-19, 30)$
\\
 & HF &  $( 0.934 , 0.975)$& $(133, 150)$& $(0.796, 0.972)$& $(109, 147)$ & $(-5.2 , 15)$ & $(-8 , 34)$
\\ 
 &  bKAGRA &  $ \cdots$& $(161 , 162)$& $(0.178, 0.960)$& $(79.4 , 143) $ & $\cdots$ & $(19, 82)$
\\ 
 & (Expected value) & 0.9425 & 130  & 0.9425 & 130 & & \\ 
%%%%%%
 $ (280 M_{\odot}, 0.0852)$ &40kg & $(0.873, 0.959)$& $(244, 296) $& $(0.836, 0.973)$& $ (240, 308)$ & $(-11, 7.5)$ & $(-51, 33)$ 
\\ 
 &  FDSQZ & $(0.934, 0.977) $&  $(274, 314)$ & $(0.887, 0.976) $& $(256, 309) $ & $(-3.6, 6.2)$ & $(-25, 41)$
\\
 & HF & $(0.974, 0.988) $ & $(311, 329)$&  $(0.951, 0.979) $ & $(291 , 325 )$ & $(-0.3, 2.9)$  & $(-9 , 31)$
\\ 
 &  bKAGRA & $(0.829 , 0.954) $ & $(229 , 297)$ & $(0.668, 0.969)$& $(211, 307)$ & $(-15, 21)$  & $(-61 , 51)$
\\
 & (Expected value) & 0.9425 & 271  & 0.9425 & 271 & & \\
\hline \hline
\end{tabular}
  \label{result2}
\end{table*}

\begin{table*}[htbp]
\caption{The same as Table~\ref{result2} but for the symmetric 68\% error regions.
  }
\begin{tabular}{l l c c c c c c}
\hline \hline
 $(M_{\rm total}, z)$ & Plan & $(a/M)_{22}$& $M_{22}/M_{\odot}$ & $(a/M)_{33}$ & $M_{33}/M_{\odot}$ & $\delta(a/M)/10^{-2}$ &$ \delta M/M_{\odot} $\\
\hline
 $ (70 M_{\odot}, 0.0113)$& 40kg & $(0.926, 0.965)$& $(60,67 ) $ & $(0.903, 0.969)  $& $ (63,74  )$ &$(-5.5 ,3.5) $ & $(-12,1 )$
\\ 
 &  FDSQZ & $(0.907, 0.939) $ &$(59,63 )$ & $( 0.762, 0.952 )$& $(54,70 ) $ & $(-6.0,14 ) $ &$(-10.0,7.1 )$ 
\\
 & HF & $ (0.902, 0.938)$ &$(59 ,63  )$ & $(0.799, 0.952)$& $(54,67) $ & $(-4.3, 11)$ &$(-7,7)$ 
\\ 
 & bKAGRA & $(0.822, 0.916)$ & $(53,61 )$ & $(0.875, 0.964)$& $(62,75 )$ & $(-15,-1.6 )$ &$( -19, -4)$
\\ 
 & (Expected value) & 0.9425 & 63  & 0.9425 & 63 & & \\
%%%%%
 $ (140 M_{\odot}, 0.0438)$ &40kg & $(0.899, 0.927)$ & $(122,129 ) $ & $(0.703, 0.942)$& $(101, 139)$ & $(-4.5,19 )$ & $(-13,25 )$
\\ 
 &  FDSQZ & $(0.911, 0.946 )$&  $(126,135 ) $ & $(0.841, 0.960)$& $(117,143 )$ &$(-6.2,7.7 )$ & $(-15,15 )$
\\
 & HF &  $( 0.943 , 0.966)$& $(137,145 )$& $(0.862, 0.960)$& $(118,142 )$ & $(-2.5 ,8.7 )$ & $(-3,24 )$
\\ 
 &  bKAGRA &  $ \cdots$& $(161 ,162 )$& $(0.436, 0.921)$& $( 90.1,135 ) $ & $\cdots$ & $( 27,71 )$
\\ 
 & (Expected value) & 0.9425 & 130  & 0.9425 & 130 & & \\ 
%%%%%%
 $ (280 M_{\odot}, 0.0852)$ &40kg & $(0.888, 0.939)$& $(251, 276) $& $(0.888, 0.963)$& $ (260,299 )$ & $(-7.8,2.2 )$ & $(-38,8 )$ 
\\ 
 &  FDSQZ & $(0.943, 0.968) $&  $(281,301 )$ & $(0.914, 0.968) $& $(266, 301) $ & $(-2.3,3.8 )$ & $(-13,28 )$ 
\\
 & HF & $(0.976, 0.985) $ & $(315, 325)$&  $(0.961, 0.977) $ & $( 298,319  )$ & $(0.1,2.0 )$  & $(-2, 23 )$ 
\\ 
 &  bKAGRA & $(0.849 , 0.923) $ & $(237,268 )$ & $(0.781, 0.945)$& $(234,295 )$ & $(-11,9.5 )$  & $(-47 ,21 )$
\\
 & (Expected value) & 0.9425 & 271  & 0.9425 & 271 & & \\
\hline \hline
\end{tabular}
  \label{result2-2}
\end{table*}

%%%%%%%%%%%%%%%%%%%%%%%%%%%%%%%%%%%%%%%%%%%%%%%%%%%%%%%%%%%
\subsection{Estimation on mass and spin under GR} 
%%%%%%%%%%%%%%%%%%%%%%%%%%%%%%%%%%%%%%%%%%%%%%%%%%%%%%%%%%%

Next, we show the 90\% error regions of black hole mass $M$ and dimensionless spin $a/M$, where $a$ is the Kerr parameter,
converted from the estimated $(f_{lm},Q_{lm})$ given in Table~\ref{result2}.
To obtain $(a,M)$ (in practice, the dimensionless quantities $(a/M, M/M_\odot)$ are used here),
we treat the numerically calculated QNMs of Kerr black holes 
by using the continued fraction method~\cite{leaver}. 
Again, since we cannot obtain a distribution of $Q_{22}$ for $M_{\rm total} = 140M_{\odot}$ with bKAGRA, the error region of $(a/M)_{22}$ of the corresponding case is not shown in Table~\ref{result2} (also in Table~\ref{result2-2}).

The quality factor has one-to-one correspondence with the dimensionless spin $a/M$, while the black hole mass depends both on the quality factor and frequency. Because of these relations, 
the error regions of $(a/M, M/M_\odot)$
slightly differ from what we expect from the estimation of $(f_{lm}, Q_{lm})$.
In practice, a smaller quality factor gives a smaller $a/M$,
and a larger quality factor and a smaller frequency derive
a smaller $M/M_\odot$.
In addition, the quality factor does not increase linearly as the dimensionless spin increases, the step of $Q$ becomes larger as $a/M$ becomes larger for a constant step of $a/M$.

In Table~\ref{result2}, first, in the case of KAGRA+40kg for the three different KAGRA+ plans,
the error regions of $(a/M, M/M_\odot)$ contain expected values for all cases although the error regions are wide for the heavier mass cases and the estimation of $(3,3)$ mode for $70M_{\odot}$ may not be reliable as mentioned before.
Second, for the KAGRA+FDSQZ case, we find that most estimations are well done although the error region of $M_{22}$ does not contain the expected value
for the case of $280M_{\odot}$.
Here, we find that the bias in the frequency estimation
affects the mass estimation a lot.
Third, estimations for the KAGRA+HF case are better when the mass of the system is lighter, i.e., the higher QNM frequency cases, while we can see that KAGRA+HF is not suitable for the larger mass case, as expected.

Table~\ref{result2-2} is the same as Table~\ref{result2}, but for the case with the 68\% error region.
The narrow error region is imitated by the large SNR case.
Even in the parameter estimation for the $(l,m)=(2,2)$ mode,
we see various deviation from the expected values.
Again, these deviation arises from physical or unphysical
bias in the estimation of $(f_{lm}, Q_{lm})$ mentioned above.

Finally, we give some comments here.
Since the numerical waveform obeys GR, the black hole parameters estimated under GR from different $(l,m)$ should coincide, i.e., $(a/M)_{22} = (a/M)_{33}$ and $M_{22} = M_{33}$.
In Table~\ref{result2}, since the error regions for $(3,3)$ mode parameters are much wider than those for $(2,2)$ mode parameters, the consistency of $(a,M)$ between two modes is achieved for all sensitivity curves and all mass scale cases.
To quantitatively see how the parameters are consistent, 
we may consider their differences for each noise realization as 
\be
\begin{split}
\delta \left(\frac{a}{M}\right) & \equiv  \left(\frac{a}{M}\right)_{22} - \left(\frac{a}{M}\right)_{33} \,,
\\
\delta M & \equiv  M_{22} - M_{33} \,.    
\end{split}
\label{eq:ER}
\ee
In Table~\ref{result2} and \ref{result2-2}, we show the 90\% and 68\% error regions of $\delta (a/M)$ and $\delta M$.
If the error regions contain 0, then we can say
that the black hole parameter for two modes is consistent.
However, the gap can be small although both modes estimate different values from GR.
Therefore, these differences are just a reference to see the consistency between two modes.
On the other hand, when we assess detector plans with GR waveforms, $\delta (a/M)$ and $\delta M$ are a useful indicator. It is found  from Table~\ref{result2-2}
that bKAGRA and KAGRA+HF give the error regions which does not contain 0.
Only in this sense, KAGRA+40kg and FDSQZ will be good candidates in the three different KAGRA+ plans.

%%%%%%%%%%%%%%%%%%%%%%%%%%%%%%%%%%%%%%%%%%%%%%%%%%%%%%%%%%%
\section{Summary} 
%%%%%%%%%%%%%%%%%%%%%%%%%%%%%%%%%%%%%%%%%%%%%%%%%%%%%%%%%%%

We have analyzed ringdown gravitational waves of binary black hole mergers by matched filtering to investigate the performances of KAGRA upgrade plans on black hole spectroscopy. 
We have used the $(l,m) = (2,2)$ and $(3,3)$ modes of gravitational waves from an RIT's NR simulation for the analysis.
Combining the two modes, Gaussian noise is added
based on noise power spectra of three different KAGRA+ plans 
(40kg, HF, and FDSQZ) and bKAGRA.
Here, we have focused only on the fundamental ($n=0$) QNM
and ignored overtone ($n \neq 0$) QNMs.
In our ringdown gravitational wave data analysis, 
damped sinusoidal templates are used for the matched filtering analysis.
Here, we obtain the best fit frequency and quality factor $(f_{lm},Q_{lm})$
for each $(l,m)$ mode as a function of the starting time of the template.
The estimated values of $(f_{lm},Q_{lm})$ are defined
by evaluating the minimum deviation of $f_{lm}$ from its mean value 
in a certain time interval after the merger time.
After estimating the parameters of $(l,m)=(2,2)$ mode, 
we remove the frequency range lower than the estimated frequency of $(2,2)$ mode 
for the matched filter integration to search for the $(l,m) = (3,3)$ mode. 
We have performed the above analysis for 100 different noise realizations for several mass scales.

We find from Tables~\ref{result1} and \ref{result1-2} that 
KAGRA+FDSQZ is the best among the three different KAGRA+ plans 
for the ringdown gravitational wave data analysis. 
In KAGRA+FDSQZ, ringdown parameters $(f_{lm},Q_{lm})$ tend to be underestimated than the expected ones,
even if the signal is observed with a sufficiently large SNR. 
This bias is natural
and will be removed when we incorporate the overtone modes
in the analysis.
The above biased parameter estimations affect
the derivation of black hole parameters $(a,M)$.
In the main text, we have assumed GR and obtain $(a,M)$ from the analyzed results of 
$(f_{lm},Q_{lm})$ (see Appendix~\ref{eikonal} for a non GR case). 
Using $(a,M)$ from each QNM, we can check the consistency of evaluated parameters
by $\delta (a/M)$ and $\delta M$ defined in Eq.~\eqref{eq:ER}
with some cautions mentioned below Eq.~\eqref{eq:ER}.
In the case of KAGRA+40kg and FDSQZ in Tables~\ref{result2} and especially \ref{result2-2}, 
the black hole parameters evaluated in GR are consistent for all total mass cases.

In our analysis, we have considered black hole binaries with total mass of $O(10^2) M_{\odot}$
where $140 M_{\odot}$ and $280 M_{\odot}$ have not yet been detected.
However, a marginal binary black hole event, IMBHC-170502 with total mass of $\sim 160 M_{\odot}$
is reported~\cite{Udall:2019wtd}, and
we can expect to detect higher significant events of binaries with such mass scale in O5.
Therefore, although our analysis focus on ideal cases,
our results may not be far from the real case.
Furthermore, a gravitational wave candidate event,
S200114f~\cite{s200114f} was found
in the coherent WaveBurst data analysis pipeline, exclusively for intermediate mass black holes, recently.
The central frequency is 64.69 Hz,
and this may be a ringdown signal
of an intermediate mass binary black hole coalescence
with total mass $\sim 300 M_\odot$.

%%%%%%%%%%%%%%%%%%%%%%%%%%%%%%%%%%%%%%%%%%%%%%%%%%%%%
\begin{acknowledgements}

This work was supported by JSPS KAKENHI Grant Number JP17H06358.
This research made use of data generated by RIT.
T.~N. is supported in part by a Grant-in-Aid for JSPS Research Fellows.
K.~S. acknowledges support from JSPS KAKENHI Grant No. JP19K14717.
H.~T. acknowledges support from JSPS KAKENHI Grant No. JP17K05437. 
H.~N. acknowledges support from JSPS KAKENHI Grant No. JP16K05347.

\end{acknowledgements}
%%%%%%%%%%%%%%%%%%%%%%%%%%%%%%%%%%%%%%%%%%%%%%%%%%%%%

\appendix
%%%%%%%%%%%%%%%%%%%%%%%%%%%%%%%%%%%%%%%%%%%%%%%%%%%%%
\section{Distributions of estimated parameters}
\label{pdf}
%%%%%%%%%%%%%%%%%%%%%%%%%%%%%%%%%%%%%%%%%%%%%%%%%%%%%

\begin{figure}[h!]
\includegraphics[scale=0.45]{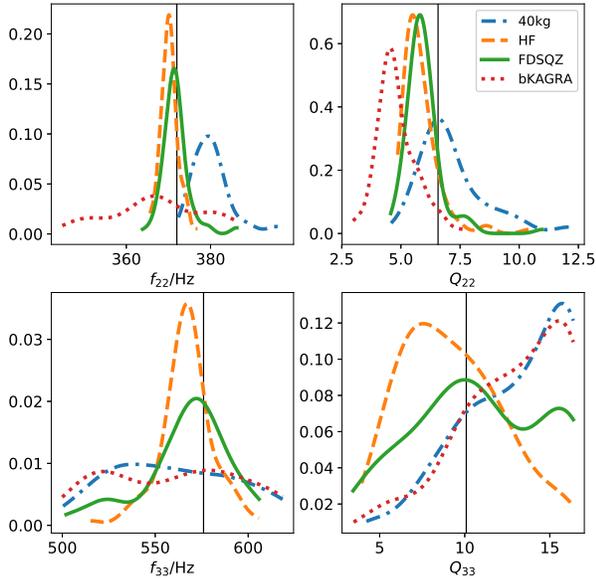}
\caption{Distributions of estimated parameters given by 100 noise realizations for three KAGRA+ plans, 40kg, HF, FDSQZ, and bKAGRA. 
The analyized waveform is scaled as $M_{\rm total} = 70 M_{\odot}$ and $z = 0.0113$. 
The expected values of $f_{22}$, $Q_{22}$, $f_{33}$, and $Q_{33}$ are 372 Hz, 6.57, 576 Hz, and 10.09,
respectively, as shown in vertical lines.}
\label{pdf1}
\end{figure}

\begin{figure}[H]
\includegraphics[scale=0.45]{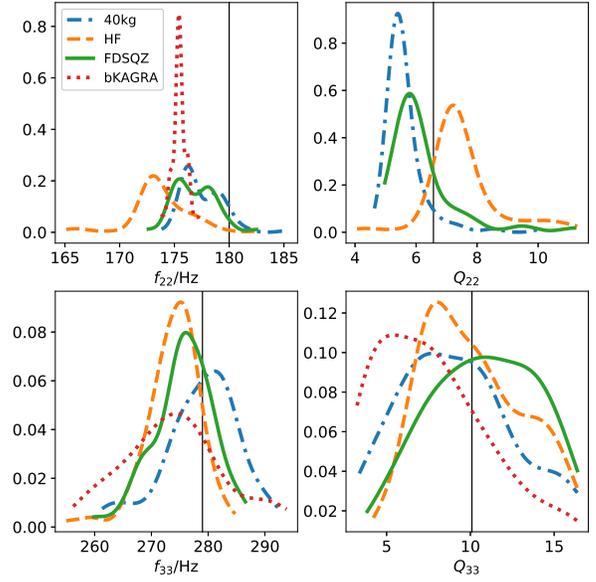}
\caption{The same as Fig.~\ref{pdf1} but for the case of $M_{\rm total} = 140 M_{\odot}$ and $z = 0.0438$.
The expected values of $f_{22}$, $Q_{22}$, $f_{33}$, and $Q_{33}$ are 180 Hz, 6.57, 279 Hz, and 10.09,
respectively, as shown in vertical lines.
In the upper-right panel,
the bKAGRA case is not shown because estimated $Q$ only takes one value.
}
\label{pdf2}
\end{figure}

\begin{figure}[H]
\includegraphics[scale=0.45]{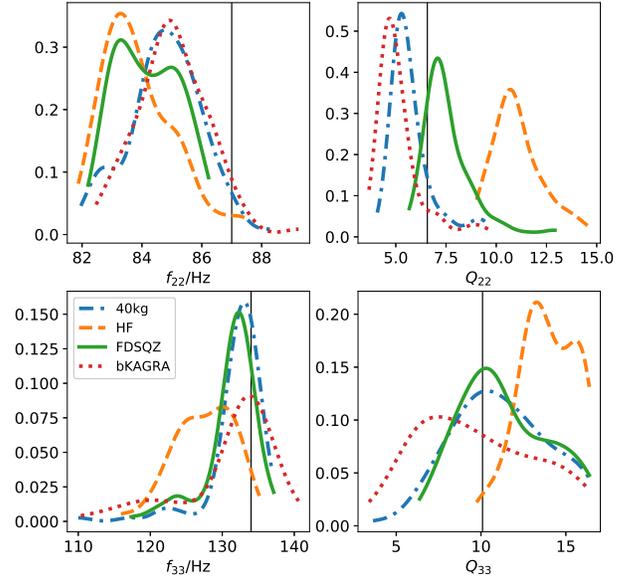}
\caption{The same as Fig.~\ref{pdf1} but for the case of $M_{\rm total} = 280 M_{\odot}$ and $z = 0.0852$.
The expected values of $f_{22}$, $Q_{22}$, $f_{33}$, and $Q_{33}$ are 87 Hz, 6.57, 134 Hz, and 10.09,
respectively, as shown in vertical lines.}
\label{pdf3}
\end{figure}

In this appendix, we show the distributions of estimated parameters $(f_{lm}, Q_{lm})$
for 100 noise realizations in Figs.~\ref{pdf1} -- \ref{pdf3}.
The distributions are normalized so that the area becomes unity.
Although our analysis is not Bayesian analysis, the distribution may be equivalent as the probability function of the parameters.
The symmetric 90\% and 68\% error regions are summarized in Tables~\ref{result1} and \ref{result1-2},
respectively.
In the upper-right panel of Fig.~\ref{pdf2},
the bKAGRA case is not shown because estimated $Q$ only takes one value, the upper end of the search region.

%%%%%%%%%%%%%%%%%%%%%%%%%%%%%%%%%%%%%%%%%%%%%%%%%%%%%
\section{Testing GR using the Johannsen-Psaltis metric}
\label{eikonal}
%%%%%%%%%%%%%%%%%%%%%%%%%%%%%%%%%%%%%%%%%%%%%%%%%%%%%

In the main text, we have analyzed ringdown gravitational waves of binary black hole mergers
and estimated black hole parameters based on GR.
In this appendix, we evaluate black hole parameters for non GR cases,
and discuss whether we can distinguish the non GR waveforms from the GR one with KAGRA+, and give any constraint.
In practice, we treat the GR waveform used in the main text and check the consistency of black hole parameters estimated in the non GR case.
Since there is no general solution for rotating black holes in modified gravity theories,
we use a post Kerr approximation proposed by Glampedakis et al.~\cite{glam}. 
They show a method to calculate QNMs
excited in the spacetime slightly deviated from the Kerr one by using the eikonal limit.
We follow this study and estimate black hole parameters from QNMs,
and evaluate the consistency of black hole parameters between two modes again.

The idea shown in Ref.~\cite{glam} is summarized in the following.
First, we consider a stationary axisymmetric non Kerr spacetime that is slightly deviated from Kerr spacetime. 
Next, QNMs in the eikonal limit, namely eikonal QNMs, 
are related to the light ring radius in this spacetime.
Therefore, we can calculate them when the metric of spacetime is given. 
Then, assuming that the offset between the true QNMs and eikonal QNMs
is the same as the Kerr one even for the non Kerr spacetime,
we obtain the full QNMs of the spacetime.

As an explicit model, we treat Johannsen-Psaltis (JP) metric~\cite{jp} for the non Kerr QNMs, which is also shown as an example in Ref.~\cite{glam}.
The JP metric is not the solution of vacuum Einstein equation,
however, we may assume the solution just as an example of non Kerr solutions. 
Each component of the JP metric $g_{\mu \nu}^{\mbox{{\scriptsize JP}}}$ is given as 
\begin{align}
g_{tt}^{\mbox{{\scriptsize JP}}} & = (1+\xi)g_{tt}^{\mbox{{\scriptsize Kerr}}} \,,
\cr
g_{t \phi}^{\mbox{{\scriptsize JP}}}  &= (1+\xi)g_{t \phi}^{\mbox{{\scriptsize Kerr}}} \,,
\cr
g_{rr}^{\mbox{{\scriptsize JP}}} & = g_{rr}^{\mbox{{\scriptsize Kerr}}}(1+\xi) \left ( 1 + \xi \f{a^2 \sin^2\theta}{\Delta}\right ) ^{-1} \,,
\cr
g_{\phi \phi }^{\mbox{{\scriptsize JP}}} & = g_{\phi \phi}^{\mbox{{\scriptsize Kerr}}}+\xi a^2 \left ( 1 +  \f{2 M r}{\Sigma}\right ) \sin^4 \theta \,,
\cr
g_{\theta \theta }^{\mbox{{\scriptsize JP}}} & = g_{\theta \theta}^{\mbox{{\scriptsize Kerr}}} \,,
\end{align}
where
\begin{align}
\Delta & = r^2 - 2M r + a^2 \,,
\cr
\Sigma & = r^2 + a^2 \cos^2 \theta \,,
\cr
\xi & = \epsilon_3 M^3 r/\Sigma^2 \,.
\end{align}
Here, $a, M$ are the Kerr parameter and black hole mass, respectively,
and $g_{\mu \nu}^{\mbox{{\scriptsize Kerr}}}$ is the Kerr metric in the Boyer-Lindquist coordinate.

\begin{table*}[htpb]
  \caption{The symmetric 90\% error regions of $(a,M)$ using the QNMs of the JP metric for $\epsilon_3 = \pm 0.1$.
  The estimation of $(f_{lm}, Q_{lm})$ from KAGRA+FDSQZ is used.
  The numerical values in the parenthesis for $(a/M)_{lm}$ and $M_{lm}/M_{\odot}$ denote the lower and upper bounds of the error region.}
\begin{tabular}{l r c c c c c c}
\hline \hline
 $(M_{\rm total}, z)$ & $\epsilon_3$ & $(a/M)_{22}$& $M_{22}/M_{\odot}$ & $(a/M)_{33}$ & $M_{33}/M_{\odot}$ & $\delta(a/M)/10^{-2}$ &$ \delta M/M_{\odot} $\\
\hline
 $ (70 M_{\odot}, 0.0113)$& 0.1 
 & $(0.863, 0.926)$ & $(57, 65)$ & $(0.433, 0.942)$ & $(44, 72)$ & $(-9.7,37)$ & $(-14,17)$
\\ 
 &  $-0.1$
 & $(0.927, 0.989)$ & $(59, 64)$ & $(0.414, 0.989)$ & $(51, 96)$  & $(-8.4,43)$ & $(-11,19)$
\\
%%%%%
 $ (140 M_{\odot}, 0.0438)$ &0.1 
 & $(0.871, 0.940)$ & $(121, 142)$ & $(0.698, 0.941)$ & $(102, 145)$ & $(-6.7,20)$ &$(-18,30)$
\\ 
 & $-0.1$ 
 & $(0.938, 0.992)$ & $(127, 145)$ & $(0.715, 0.991)$ & $(104, 147)$ &$(-5.8,23)$ & $(-15,33)$
\\
%%%%%%
 $ (280 M_{\odot}, 0.0852)$ &0.1 
 & $(0.901, 0.950)$ & $(267, 317)$ & $(0.859, 0.948)$ & $(252, 306)$ &$(-3.7,6.3)$ & $(-28,45)$
\\ 
 & $ -0.1$ 
 & $(0.974, 0.996)$ & $(288, 312)$ & $(0.915, 0.996)$ & $(263, 306)$ & $(-2.1,6.6)$&$(-8,40)$
\\
\hline \hline
\end{tabular}
  \label{postkerr2}
\end{table*}

The parameter $\epsilon_3$ shows the deviation from the Kerr metric,
and the quadrupole moment of the JP metric is given as
\be
Q^{\mbox{{\scriptsize JP}}} = \f{J^2}{M} - \epsilon_3 M^3 \,,
\ee
where $J = a M$.
This means positive (negative) $\epsilon_3$ corresponds to a prolate (oblate) deformation.

Here, we briefly review the derivation of QNMs of the JP metric in the eikonal limit, $(\sigma_R, \sigma_I)$.
To obtain the light ring radius $r_0$ and the impact parameter $b$, 
we need to solve $V_{\rm eff}(r_0)=0$ and $V_{\rm eff}^{\prime}(r_0)=0$, 
where $V_{\rm eff}$ is an effective potential of the radial motion for null geodesics.
The impact parameter is the inverse of the angular frequency at the light ring $\Omega$, 
which gives the real part of the QNMs in the eikonal limit.
These equations lead to
\begin{align}
& r_0^6+(r_0^3+M^3 \epsilon_3 ) \left [ 2M (a-b)^2 + r_0 (a^2-b^2) \right ] =0 \,,
\label{rb1} \\
& 2 r_0^3  \left [r_0^3 - M(a-b) ^2 \right ]\nn \\
&\qquad \>  \>  \, 
- \epsilon_3 M^3 \left [8M(a-b)^2 + 3 r_0 (a^2 - b^2 ) \right ] = 0 \,,
\label{rb2}
\end{align}
for the JP metric case.
Alternatively, we may use a little simple equation,
\begin{align}
& 2 r_0^3  \left [r_0 (a^2-b^2) + 3M(a-b) ^2 \right ]\nn \\
&\qquad \>  \>  \, 
+ \epsilon_3 M^3 \left [12M(a-b)^2 + 5 r_0 (a^2 - b^2 ) \right ] = 0 \,.
\end{align}
Next, we need the Lyapunov exponent, which corresponds to the convergence/divergence 
of the noncircular orbit from the light ring radius,
and it gives the imaginary part of the QNMs in the eikonal limit.
From Eqs.~(44) and (47) in Ref.~\cite{glam}, we have
\be
\begin{split}
& \gamma(r_0,b) = \gamma_K(r_0,b) \\
& \quad \times \left [  1 + 2 \f {\epsilon_3 M^3 }{r_0^3} \left (  f_1 - f_2 \right ) + \f {\epsilon_3^2 M^6 f_2 }{r_0^6} \left ( 2 +f_2 - 4f_1   \right ) \right ] \,, 
\end{split}
\ee
where
\be
\gamma_K(r_0,b) = \f {(r_0^2-2 M r_0 + a^2)}{b \left[ 2M (a -b) + b r_0 \right]} 
\sqrt {   \f{2M(a-b)^2 + r_0^3}{r_0^3} } \,,
\label{gammak}
\ee
and
\be
\begin{split}
f_1 & = \f {10M(a-b)^2+3r_0(a^2 - b^2)}{r_0^3+2M(a-b)^2} \,, \cr
f_2 & = \f{r_0(r_0-2M)}{r_0^2 -2 M r_0 + a^2} \,.
\end{split}
\ee
Equation~\eqref{gammak} is the Lyapunov exponent for Kerr spacetime, which becomes
\be
\gamma_K(r_0,b) = \f {2\sqrt{3 M} \Delta }{b r_0^{3/2} (r_0-M)} \,.
\ee
Here, we have used the relation between $r_0$ and $b$ from Eqs.~\eqref{rb1} and \eqref{rb2} with $\epsilon_3=0$.
By numerically solving Eqs.~\eqref{rb1} and \eqref{rb2} for $r_0$ and $b$, we obtain 
\be
\begin{split}
\sigma_R &= m \Omega = \frac{m}{b} \,, \\
\sigma_I &= -\frac{|\gamma(r_0,b)|}{2} \,.
\end{split}
\ee
Then, using the offset between the exact QNMs and those in the eikonal limit
for the Kerr black hole $\beta_K$, we have 
\be
\begin{split}
\omega_R^{\rm JP} &= \sigma_R + {\rm Re} (\beta_K) \,, \\
\omega_I^{\rm JP} &= \sigma_I + {\rm Im}( \beta_K) \,.
\end{split}
\ee
Note that the dependence of $(l,m)$ is included in $\beta_K$.

Using $(f_{lm}, Q_{lm})$ shown in Table~\ref{result1},
we calculate corresponding black hole parameters for the JP metric for each mode
by applying the method
explained above to obtain QNMs for the JP metric.
Here, we also choose the estimation of $(f_{lm}, Q_{lm})$ from the KAGRA+FDSQZ case,
since we are considering that this configuration is the most suitable for black hole spectroscopy.
In this JP case, the QNMs are determined by three parameters $(a,M,\epsilon_3)$. 
We expect to see significant differences of black hole parameters between two modes for each $\epsilon_3$,
because an NR waveform assuming GR
has been used in this study.

\begin{figure}[!t]
\includegraphics[scale=0.55]{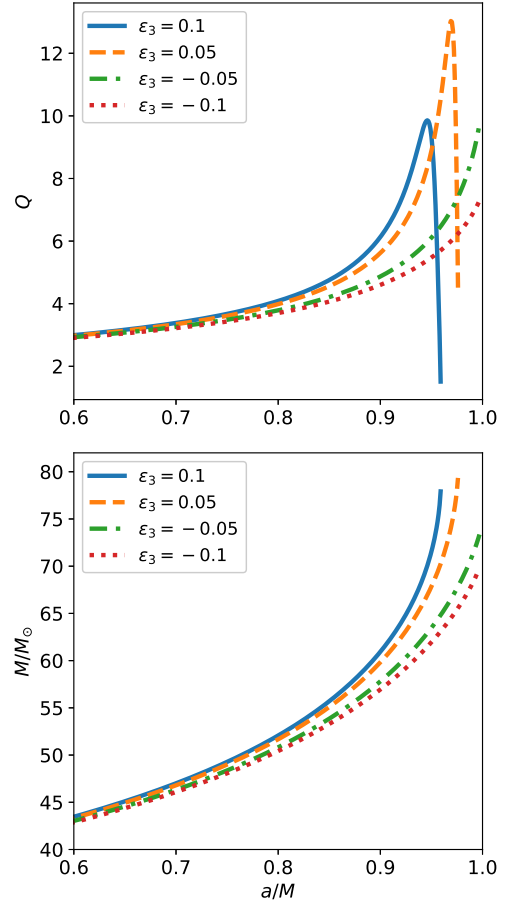}
\caption{Top: The behavior of the quality factor against the Kerr parameter in the case of $\epsilon_3= \pm 0.1$ and $\pm0.05$ for the $(l,m)=(2,2)$ QNM.  Bottom:The behavior of the estimated mass against the Kerr parameter when $f = 370$Hz in the case of $\epsilon_3= \pm 0.1$ and $\pm0.05$ for the $(l,m)=(2,2)$ QNM. 
}
\label{pk1}
\end{figure}

In Refs.~\cite{jp} and \cite{glam}, to confirm the constraint from gravitational wave observation,
they have considered $|\epsilon_3| \le 10$.
In the top panel in Fig.~\ref{pk1}, we show the behavior of the quality factor against the Kerr parameter
for $\epsilon_3 = \pm 0.1$ and $\pm0.05$.
We find in the case of $\epsilon_3 > 0$ that the relation between $Q$ and $a/M$ is not
given by a single valued function.
The relation between the mass and Kerr parameter with a fixed frequency
is presented in the bottom panel in Fig.~\ref{pk1}.
For positive $\epsilon_3$, the mass diverges at some $a/M$
in the range of $a/M \leq 1$.
These strange behaviors may be related to the modification of extreme limit of the Kerr parameter,
but the analysis is beyond the scope of this paper (see Ref.~\cite{jp} where, for example, the light ring radius is the same as that of the event horizon around $a/M=0.697$ in the case of $\epsilon_3=2$
and the horizon does not close beyond this value of the Kerr parameter).
Anyway, we use only the part of increasing function with respect to $a/M$ for $Q$ in our analysis.
We should also note that in the restricted parameter region of $a/M \leq 1$,
the maximum value of the quality factor decreases when $|\epsilon_3|$ becomes larger.
The expected $Q_{22}$ in the main text is $6.57$.
Therefore, we consider only for the $\epsilon_3 = \pm 0.1$ cases. 

The symmetric 90\% regions of black hole parameters for each mode for $\epsilon_3= \pm 0.1$ are shown in Table~\ref{postkerr2}.
As mentioned above, the quality factor saturates for the JP metric, we just adjust the maximum value of $a/M$ when the estimated $Q$ exceeds the theoretical value of $Q$.
Checking the 90\% error regions of $\delta (a/M)$ and $\delta M$,
we find that the error regions defined in Eq.~\eqref{eq:ER} contain 0.
Therefore, in the current situation,
the black hole parameters for two modes is consistent for $\epsilon_3= \pm 0.1$,
and we cannot refute the JP metric with $\epsilon_3= \pm 0.1$.

%%%%%%%%%%%%%%%%%%%%%%%%%%%%%%%%%%%%%%%%%%%%%%%%%%%%%%%%%%%

\end{document}